# Spin transport and polarization properties of manganese-doped dual-guanine molecule

Hamidreza Simchi [a,b,]*, Mahdi Esmaeilzadeh [a], Hossein Mazidabadi [b], and Mohammad Nourozi [b]

[a] Department of Physics, Iran University of Science and Technology, Narmak, Tehran 16844, Iran

[b] Semiconductor Technology Center, Tehran, Iran

We study the spin transport and polarization properties of manganese-doped dual-guanine molecules connected to graphene leads using non-equilibrium Green's function method. It is shown that a manganese doped dual-guanine molecule is a biological semiconductor and behaves as a prefect spin filter. We show that this semiconductor can behave as a spin switch when the Rashba spin-orbit interaction is considered. In addition, it is shown that, a large conductance is observed due to the Fano-Kondo-Rashba resonance effect.

**Keywords:** Guanine molecule, Spintronics, Rashba spin-orbit interaction, Spin Polarization

*Corresponding author at: Department of Physics, Iran University of Science and Technology, Narmak, Tehran 16844, Iran . E-mail address: simchi@iust.ac.ir

## I. Introduction

Nowadays, technologists search different procedures for creating the high circuit integration density on a single chip by scaling down the electronic devices. Between different possible methods for doing so, low-cost and easy-to-process methods are more attractive. Molecular electronics, which consists of the use of molecules to realize electrically conducting structures, is one of the main approaches considered. Organic field effect transistors (OFETs) [1], organic light-emitting devices (OLEDs) [2], and organic solar cells[3] are a few examples of molecular electronics. In addition, DNA molecules are attracting attention for the construction of nanometer scale devices because they have self-assembling capabilities and sequencing-specific recognition properties[4]. The DNA molecule includes four nucleobases called adenine, guanine, thymine, and cytosine. Among other nucleobases, guanine has the lowest ionization potential. Therefore, the charge motion along a sequence of nucleobases are mediated by guanine.[5,6]

Several assemblies of guanine molecule have been investigated by the method of first-principles calculations, and the role of $\pi$-$\pi$ coupling and band transport have been explained.[7] Beleznay et al. [8] have derived the correct expression of charge mobility in the one-dimensional guanine stack in the deformation potential approximation in detail. The temperature dependence and anisotropy of charge-carrier transport in guanine-based materials have been studied using a guanine crystal as a model system.[9] Oetzal et al.[10] have considered one-dimensional structures of guanine and melanin and studied the electronic and transport properties of these structures. We have studied the spin transport and polarization properties in zinc and manganese doped adenine in the presence and the absence of an applied external electrical field.[11]

In this paper, we consider an asymmetric dual-guanine molecule connected to graphene leads for studying the spin switching and spin filtering effects in a biological molecule. Using non-equilibrium Green's function method, we calculate the spin-dependent conductance and spin polarization. We show that manganese-doped dual a molecule is a biological semiconductor and behaves as a prefect spin filter. By adding the Rashba spin orbit interaction (RSOI) to Hamiltonian, it is shown that this semiconductor can behave as a spin switch. Finally, we show that a large-conductance is seen due to the Fano-Kondo-Rashba resonance effect. In section II, we explain the calculation method and in section III the results and discussion are provided. Finally, in section IV the summary is presented.

## II. Calculation method

Fig. 1 shows the manganese-doped dual guanine molecule connected to the graphene leads. We optimize all structures using Gaussian code.[12] Of course, one can use existing optimized parameters of single guanine molecule[13,14] and then optimize dual guanine molecule connected to graphene leads.[11] For calculating the conductance at zero bias regimes, we should calculate Fock and overlap matrices. Fock and overlap matrices can be calculated by using the non-standard commands of Gaussian code i.e., iop(5/33=3) and iop(3/33=1), respectively.[12] The molecular orbitals of each atom are expanded in terms of Gaussian basis functions. Table 1 to 3 show the molecular orbital expansion in terms of 3-21G basis function.[12] Since the Gaussian basis functions are not orthogonal, we define an effective Hamiltonian as $H_{eff} = (I - Overlap) \times E + Fock$ where $E$ is electron energy and $I$ is unit matrix.[15,16] The Green's function can be calculated using

$$\mathbf{G} = ((E + i\eta)I - H_{eff} - \Sigma_L - \Sigma_R)^{-1} \quad (1)$$

where $\eta$ is an infinitesimal number and $\Sigma_{L(R)}$ is the self energy of left (right) lead. The surface Green's functions and self energy of left and right leads are found using Lopez algorithm.[15-17] The zero bias conductance is given by [15,16]

$$G = (e^2/h) Real\left[Trace(\Gamma_L \mathbf{G} \; \Gamma_R \mathbf{G}^{\dagger})\right] \tag{2}$$

where $\Gamma_{L(R)}$ is the coupling matrix of the left (right) lead, $-e$ is the electron charge and $h$ is the Plank constant.

The Rashba spin-orbit interaction Hamiltonian is given by:

$$H_{SO}^{Rashba} = \lambda_{Rashba} \begin{pmatrix} 0 & \partial_y + i\,\partial_x \\ \partial_y - i\,\partial_x & 0 \end{pmatrix} \tag{3}$$

where $\lambda_{Rashba}$ is Rashba spin-orbit strength. We assume $\lambda_{Rashba}$ is equal to 0.065 eV Ang., 0.13 eV Ang., and 0.5 eV Ang. (i.e., 0.019, 0.0048, and 0.0024 a.u. Ang.), respectively which is 2.5, 10 and 20 times smaller than that of Ref. [18]. The total Hamiltonian is given by :

$$H = \begin{pmatrix} H_0 & 0 \\ 0 & H_0 \end{pmatrix} + \lambda_{Rashba} \begin{pmatrix} 0 & \partial_y + i\,\partial_x \\ \partial_y - i\,\partial_x & 0 \end{pmatrix} \tag{4}$$

where $H_0$ is the Hamiltonian without Rashba spin-orbit interaction. The elements of Rashba Hamiltonian can be calculated analytically by using the basis functions of table 1-3.[15] Using Eq.1, we calculate the Green's function of channel in presence of Rashba spin-orbit interaction. The Green's function matrix is decomposed into four parts as shown below:

$$\mathbf{G} = \begin{pmatrix} \mathbf{G}^{\uparrow\uparrow} & \mathbf{G}^{\uparrow\downarrow} \\ \mathbf{G}^{\downarrow\uparrow} & \mathbf{G}^{\downarrow\downarrow} \end{pmatrix} \tag{5}$$

where $\uparrow$ means spin up and $\downarrow$ means spin down. By definition, $G^{\uparrow} = G^{\uparrow\uparrow} + G^{\uparrow\downarrow}$ , $G^{\downarrow} = G^{\downarrow\downarrow} + G^{\downarrow\uparrow}$, and spin polarization is $P_S = (G^{\downarrow} - G^{\uparrow})/(G^{\downarrow} + G^{\uparrow})$.

## III. Results and discussion

Fig. 2 shows the conductance of non-doped and manganese-doped dual guanine molecule versus electron energy. As Fig.2 demonstrates, for $-0.01 \prec E \prec 0.01$eV the conductance of non-doped molecule is zero and therefore the transmission gap is equal to 0.02 eV. For $-0.09 \prec E \prec 0.06$eV the conductance of mangenase-doped molecule is zero and therefore the doped molecule is a biological semiconductor with transmission gap equal to 0.15 eV. On the other hand, the doped molecule includes 569 electrons and it means that its multiplicity ($2 \times S + 1$), which $S$ is spin, is equal to two. Therefore it is expected that the Hamiltonian of spin up electrons to be different from that of spin down electrons. By using Eq.(1) and Eq.(2), it means that the conductance of spin up electrons differs from spin down electrons (see Fig.2).

Fig. 3 shows spin-dependent conductance versus electron energy when RSOI is ( is not) considered and $-1 \leq (E - E_F) \leq 1$eV. As Fig.3 (b) demonstrates, the transmission band gap is equal to 0.82 eV when RSOI strength is equal 0.065 eV Ang.. Therefore, the transmission band gap increases. However, as Figs.3 (c) and (d) show, by more increasing the RSOI strength (i.e., λ=0.13 and 0.5 eV Ang.) new conductance peaks are seen in the transmission gap (also, see table 4). As table 4 shows, the edge of LUMO region is placed at 0.06 eV for both λ=0.13 and 0.5 eV Ang. while, the edge of HOMO region is placed at -0.04 and -0.22 eV for λ=0.13 and 0.5 eV Ang., respectively. Therefore, the transmission gap decreases (increases) when λ=0.13 (λ=0.5) eV Ang. respect to the λ=0.0 eV Ang. and it always decreases respect to the λ=0.065 eV Ang. Figs.3 (b), (c), and (d) show a large conductance for some energies. The Rashba spin–orbit interaction can be due to a structure inversion asymmetry[19]. Fano effect can be attributed to the interference between localized energy levels and the continuum.[20,21] Serra and Sanchez have shown that, Fano line shapes appear in the conductance of a semiconductor quantum wire when a local Rashba spin–orbit interaction is present.[21] Orellana et al. have shown that, the Rashba spin–orbit coupling opens a new tunneling channel through a quantum dot (QD). [22] Also, they have shown that, as the Rashba spin–orbit strength varies, the conductance can pass from prefect transmission to prefect reflection.[22] Stefanski[23] has considered a QD coupled to spin-polarized metallic leads. He has shown that, the coulomb repulsion between the channels modifies the width and shape of the Fano resonance as compared to the non-interacting case. [23] We have shown that, the RSOI can open new transmission channel in a Benzene molecule connected to graphene leads.[15] Gores et al. have introduced a bridge model for studying the transport properties of a QD strongly coupled with electrodes.[24] In their model, the electronic wave can travel either directly between the electrodes or through impurity states. An electron is localized on the impurity state and in consequence, the Fano-Kondo resonance is expected in which spin-flip process occurs as well.[24] The atomic number of mangenase is 25. It means that, the element includes a $3d^5$ orbital with spin-textured as ↑↑↑↑↑ in d-orbital (↑ means spin up). Therefore, there are five uncompensated spins. We can assume that there are two main transport channels. First channel is through the manganese molecule and second one is through remainder part of doped guanine molecule. Under this status, the RSOI is applied. Therefore, we not only encounter the Fano-Rashba, but also the Fano-Kondo resonance effect. Therefore, the large conductance is created by Fano-Kondo-Rashba resonance effect.

Spin polarization is another important parameter. Fig.4 shows the spin polarization versus electron energy when RSOI is (is not) considered and $-1 \le (E - E_F) \le 1$ eV. As Fig.2 shows, the conductance of spin down (spin up) electrons is equal to 0.75$G_0$ (0.0$G_0$) when E= -0.11 eV and RSOI strength is equal to zero. Therefore, the molecule behaves as a prefect spin filter with efficiency %75. Fig. 4 shows the spin polarization versus electron energy when RSOI is (is not) considered. It should be noted that, inside the transmission gap the conductance is very small but since the spin polarization is calculated using the formula $P_S = (G^{\downarrow} - G^{\uparrow})/(G^{\downarrow} + G^{\uparrow})$, $P_S$ may even be equal to $\pm 1$ inside the transmission gap (see Fig.4). But, we ignore the spin polarization inside the transmission gap. Figs. 3(b) shows that the conductance is equal to zero for E= - 0.11 eV when spin-orbit coupling strength is equal to 0.065 eV Ang.. Therefore, although as Fig. 4(b) shows the spin polarization is not zero for E= -0.11 eV, one can still close the spin switch by applying the RSOI. In the other words, the molecule can behave as a spin switch by applying the RSOI. As Fig. 4 shows, the spin polarization is placed at range $-0.5 \le P_S \le 0.5$ when $(E - E_F) \le -0.1$ and $(E - E_F) \ge 0.06$ eV.

## IV. Summary

We have studied the spin transport and polarization properties of an asymmetric manganese-doped dual guanine molecule connected to graphene leads. We optimized all structures using Gaussian code. For calculating the spin conductance and spin polarization versus electron energy, we have used non-equilibrium Green's function method. We have shown that the transmission gap is equal to 0.02 eV and 0.15 eV in non-doped and manganese-doped molecule, respectively and therefore the doped molecule is a biological semiconductor. It has been shown that, the doped molecule behaves as a prefect spin filter with efficiency %75. By adding the Rashba spin orbit interaction to Hamiltonian, we have shown that the doped molecule can behave as a spin switch. It has been shown that, a large conductance is seen due to the Fano-Kondo-Rashba resonance effect.

**Figure caption**

**Fig. 1 (**Color online) Manganese doped dual-guanine molecule connected to graphene leads. Carbon atoms are shown in gray color, Nitrogen atoms in blue color, Oxygen **atoms in red** color, Manganese atom in green color, and Hydrogen atoms in pink color.

**Fig. 2** (Color online) Zero bias conductance versus electron energy of non-doped and Manganese doped dual guanine molecule connected to graphene leads.

**Fig.3** (Color online) Conductance versus electron energy when (a) $\lambda$=0.0 eV, (b) $\lambda$=0.065 eV Ang., (c) $\lambda$=0.13 eV, and (d) $\lambda$=0.5 eV.

**Fig. 4** (Color online) Spin polarization versus electron energy (a) without Rashba SOI and (b) with Rashba SOI

**Table caption**

**Table 1-** Molecular orbital expansion of carbon, nitrogen , and hydrogen atoms in terms of 3-21G Gaussian Basis functions

**Table 2-** Molecular orbital expansion of oxygen atoms in terms of 3-21G Gaussian Basis functions

**Table 3-** Molecular orbital expansion of manganese atoms in terms of 3-21G Gaussian Basis functions

**Table 4-** Conductance of molecule in presence and absent of Rashba SOI at special points

**Table 1- Molecular orbital expansion of carbon, nitrogen , and hydrogen atoms in terms of 3-21G Gaussian Basis functions**

| Atom | Expansion | Atom | Expansion |
|---|---|---|---|
| C | **1S:** $0.062\exp(-172.256r^2)+0.395\exp(-25.911r^2)+0.701\exp(-5.533r^2)$<br>**2S:** $-0.396\exp(-3.665r^2)+1.216\exp(-0.77r^2)$<br>**2P$_{x,y,z}$:** $0.236(x, y, z)\exp(-3.66r^2)+0.861(x, y, z)\exp(-0.77r^2)$<br>**3S:** $\exp(-0.196r^2)$<br>**3P$_{x,y,z}$:** $(x, y, z)\exp(-0.196r^2)$ | N | **1S:** $0.0599\exp(-242.766r^2)+0.353\exp(-36.485r^2)+0.707\exp(-7.814r^2)$<br>**2S:** $-0.413\exp(-5.425r^2)+1.224\exp(-1.149r^2)$<br>**2P$_{x,y,z}$:** $0.238(x, y, z)\exp(-5.425r^2)+0.859(x, y, z)\exp(-1.149r^2)$<br>**3S:** $\exp(-0.283r^2)$<br>**3P$_{x,y,z}$:** $(x, y, z)\exp(-0.283r^2)$ |
| H | **1S:** $0.156\exp(-.447r^2)+0.905\exp(-0.824r^2)$<br>**2S:** $\exp(-0.183r^2)$ | | |

**Table 2- Molecular orbital expansion of oxygen atoms in terms of 3-21G Gaussian Basis functions**

| | |
|---|---|
| **O** | **1S:** $0.0592\exp(-322.037r^2)+0.351\exp(-48.431r^2)+0.708\exp(-10.421r^2)$<br>**2S:** $-0.404\exp(-7.403r^2)+1.221\exp(-1.576r^2)$<br>**2P$_{x,y,z}$:** $0.244(x, y, z)\exp(-7.403r^2)+0.854(x, y, z)\exp(-1.576r^2)$<br>**3S:** $\exp(-0.374r^2)$<br>**3P$_{x,y,z}$:** $(x, y, z)\exp(-0.374r^2)$ |

**Table 3- Molecular orbital expansion of manganese atoms in terms of 3-21G Gaussian Basis functions**

| Atom | Expansion |
|---|---|
| Mn | **1S:** $0.0637\exp(-3041.686r^2)+0.377\exp(-460.090r^2)+0.681\exp(-100.596r^2)$<br>**2S:** $-0.110\exp(-131.767r^2)+0.0981\exp(-28.569r^2)+0.958\exp(-8.660r^2)$<br>**2P$_{x,y,z}$:** $0.140(x, y, z) \exp(-131.767r^2)+0.558(x, y, z) \exp(-28.569r^2)+0.471(x, y, z) \exp(-8.660r^2)$<br>**3S:** $-0.292\exp(-8.569r^2)+0.344\exp(-2.768r^2)+0.845\exp(-0.887r^2)$<br>**3P$_{x,y,z}$:** $0.0242(x, y, z) \exp(-8.569r^2)+0.469(x, y, z) \exp(-2.768r^2)+0.607(x, y, z) \exp(-0.887r^2)$<br>**4(XX,YY,ZZ):** $0.265(x^2, y^2, z^2) \exp(-11.069r^2)+0.852(x^2, y^2, z^2) \exp(-2.731r^2)$<br>**4(XY, XZ, YZ):** $0.265(xy, xz, yz) \exp(-11.069r^2)+0.852(xy, xz, yz) \exp(-2.731r^2)$<br>**5(XX, YY, ZZ):** $(x^2, y^2,z^2) \exp(-0.669r^2)$<br>**5(XY, XZ, YZ):** $(xy, xz, yz) \exp(-0.669r^2)$<br>**6S:** $-0.23\exp(-0.767r^2)+1.09\exp(-0.092r^2)$<br>**6P$_{x,y,z}$:** $0.0031(x, y, z) \exp(-0.767r^2)+0.9999(x, y, z) \exp(-0.092r^2)$<br>**7S:** $\exp(-0.0322r^2)$<br>**7P$_{x,y,z}$:** $(x, y, z) \exp(-0.0322r^2)$ |

**Table 4- Conductance of molecule in presence and absent of Rashba SOI at special points**

| λ-values (eV Ang.) | $(E - E_F)$ (eV) | $G(G_0)$ | Spin direction |
|---|---|---|---|
| 0.0 | -0.09 | 0.0 | ↑ |
| | | 0.0 | ↓ |
| | 0.06 | 0.0 | ↑ |
| | | 0.0 | ↓ |
| 0.065 | -0.09 | 0.0 | ↑ |
| | | 0.0 | ↓ |
| | 0.06 | 0.0 | ↑ |
| | | 0.0 | ↓ |
| 0.13 | -0.04 | 38.39 | ↑ |
| | | 23.68 | ↓ |
| | 0.06 | 0.0 | ↑ |
| | | 14.73 | ↓ |
| 0.5 | -0.22 | 0.0 | ↑ |
| | | 11.21 | ↓ |
| | 0.06 | 163.4 | ↑ |
| | | 158.46 | ↓ |

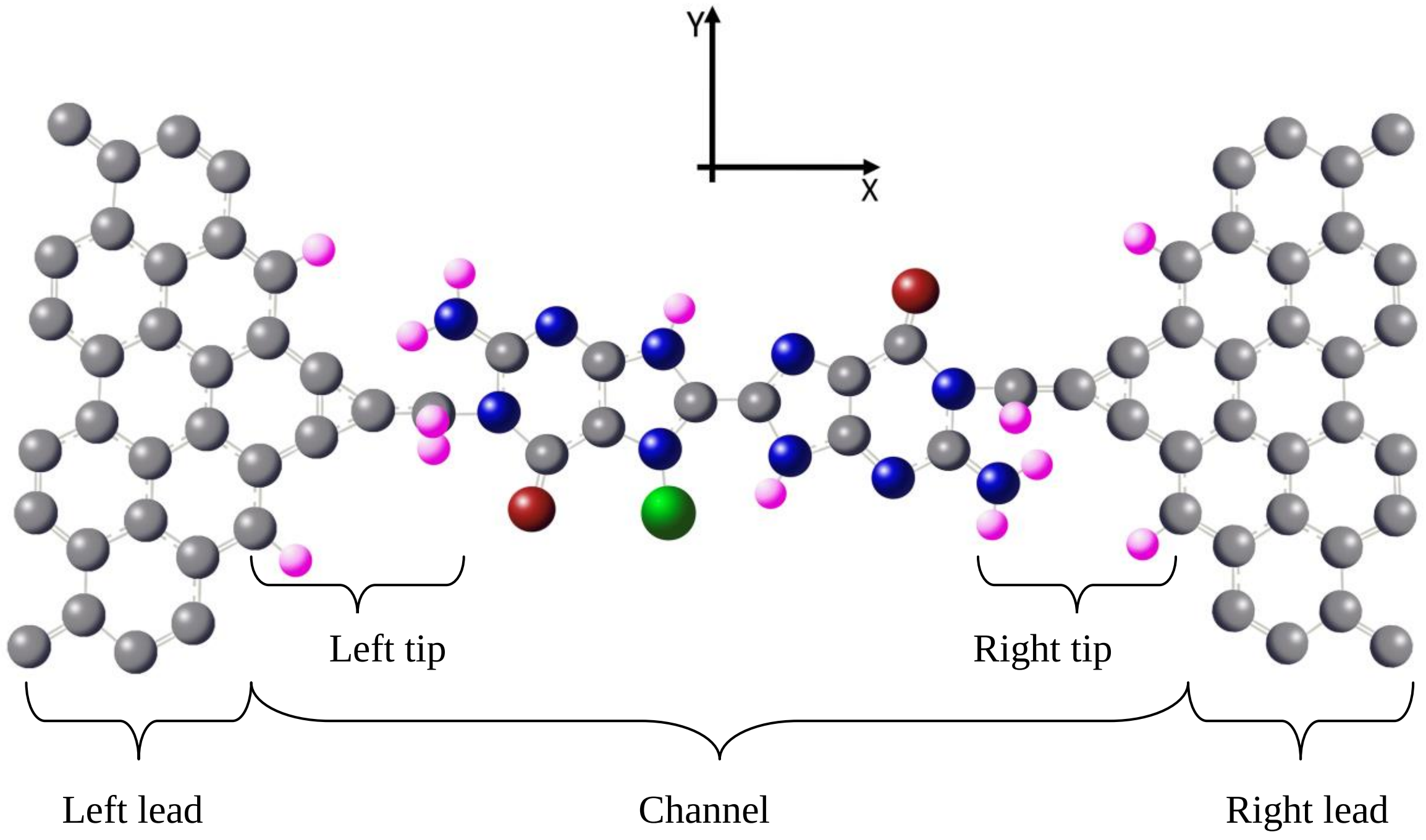

**Fig. 1 (Color online) Manganese doped dual-guanine molecule connected to graphene leads. Carbon atoms are shown in gray color, Nitrogen atoms in blue color, Oxygen atoms in red color, Manganese atom in green color, and Hydrogen atoms in pink color.**

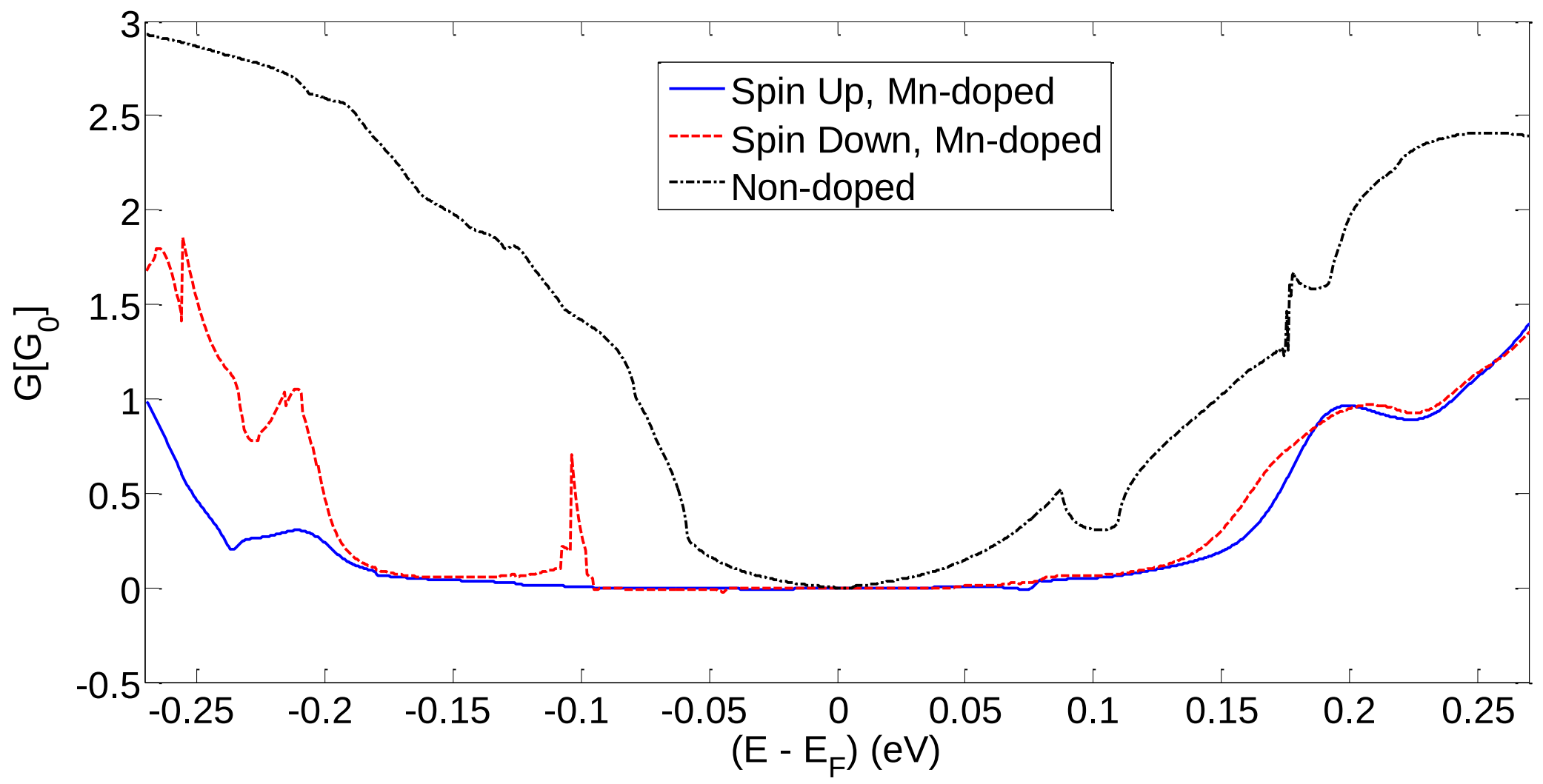

**Fig. 2 (Color online) Zero bias conductance versus electron energy of non-doped and Manganese doped dual guanine molecule connected to graphene leads.**

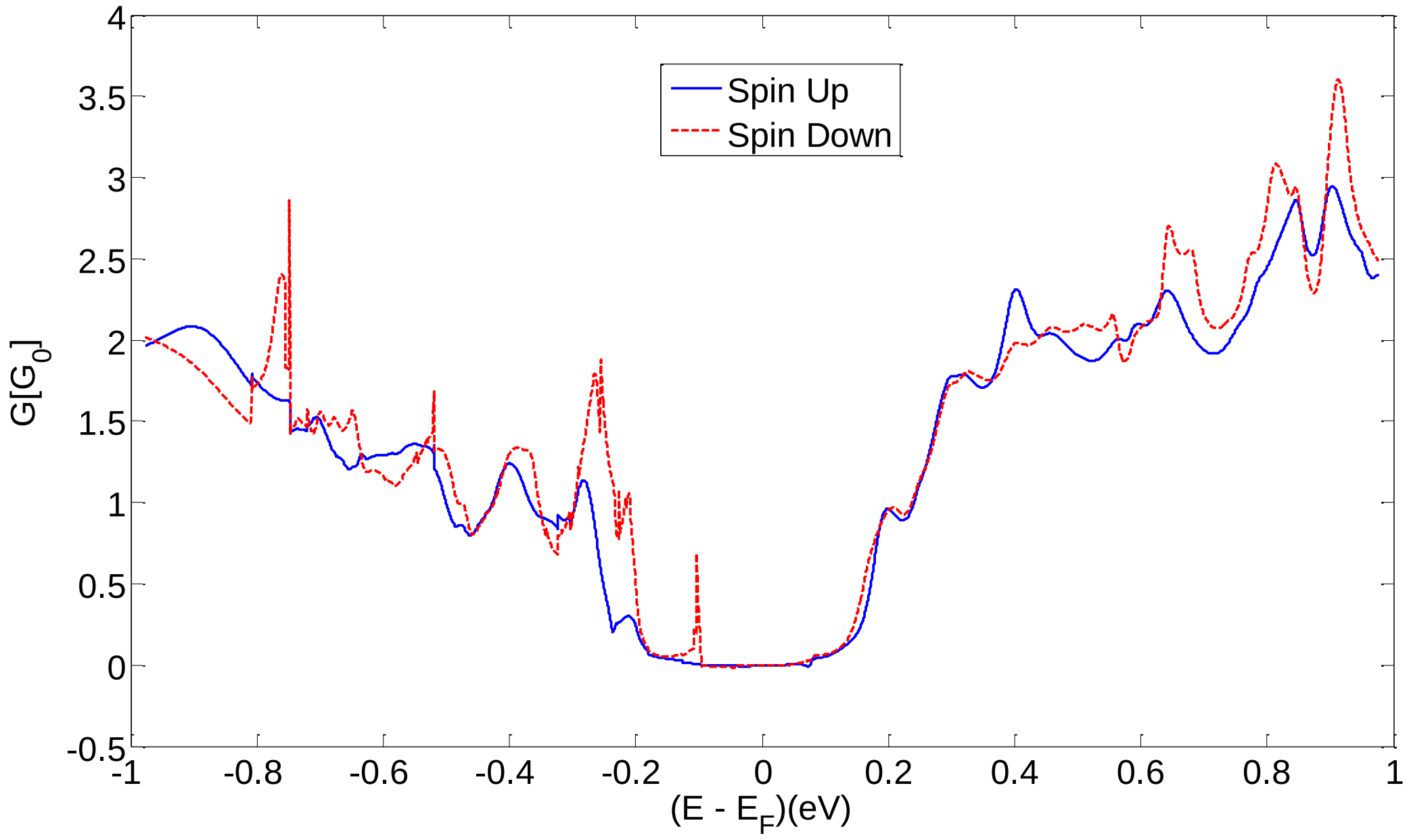

(a)

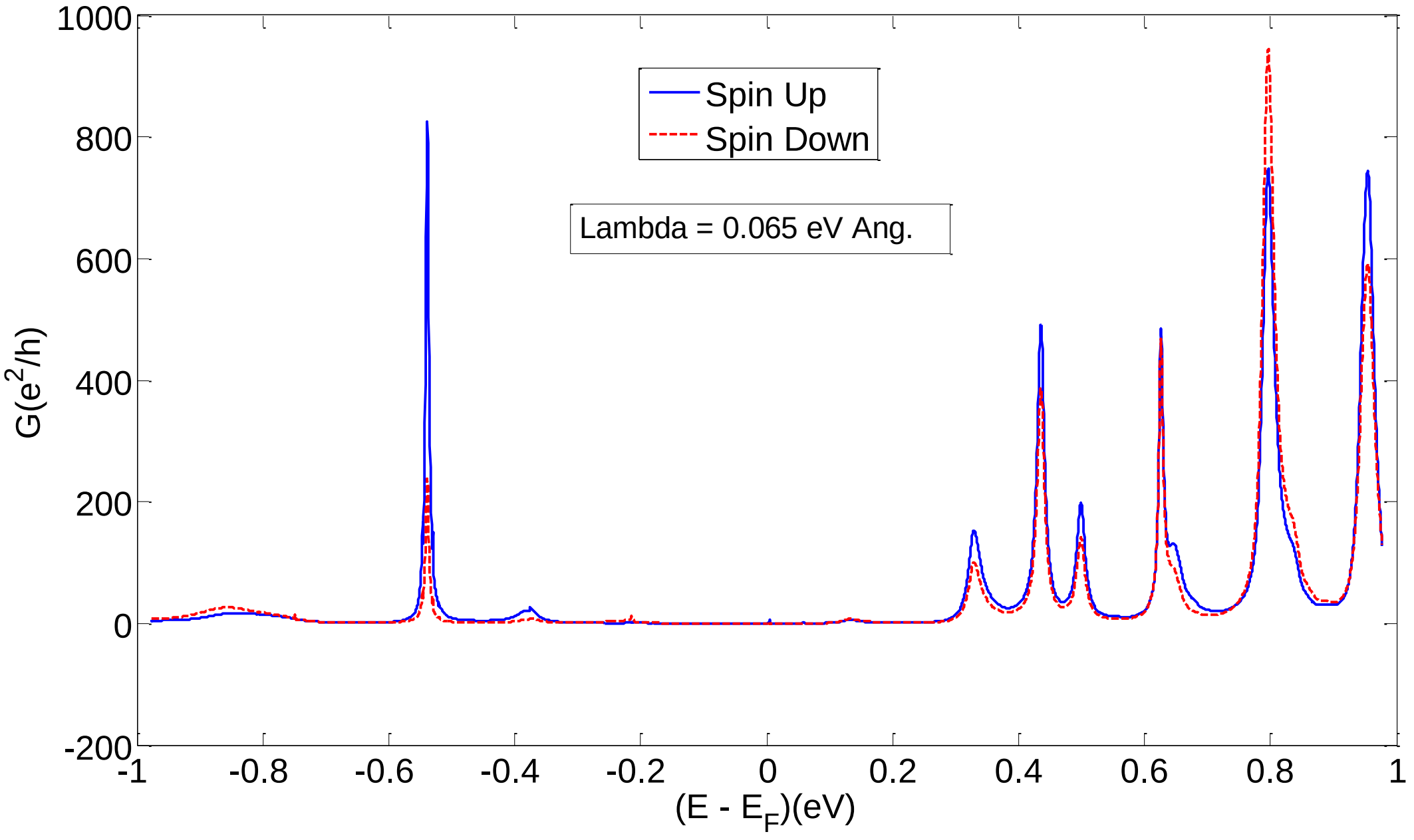

(b)

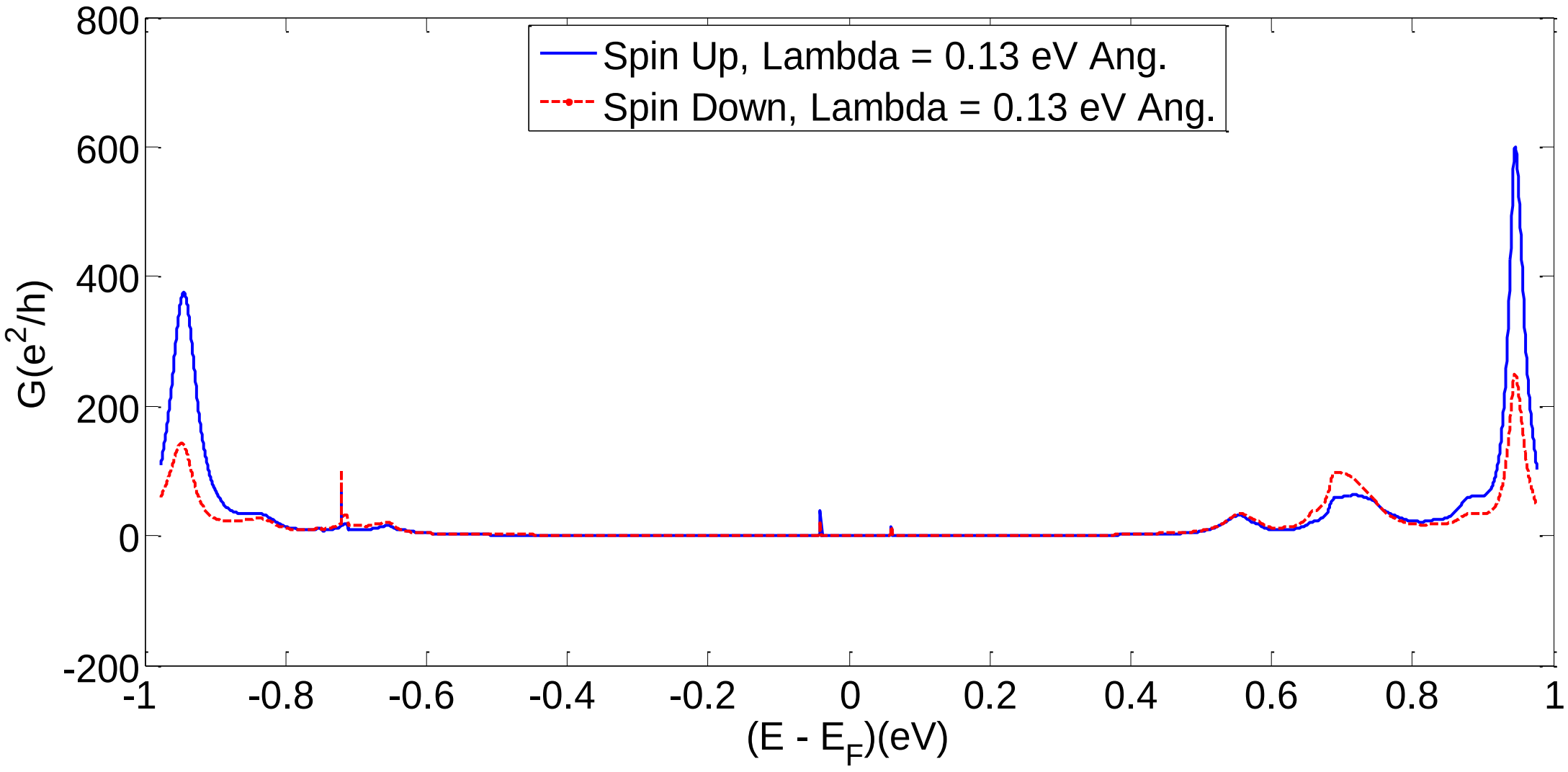

(C)

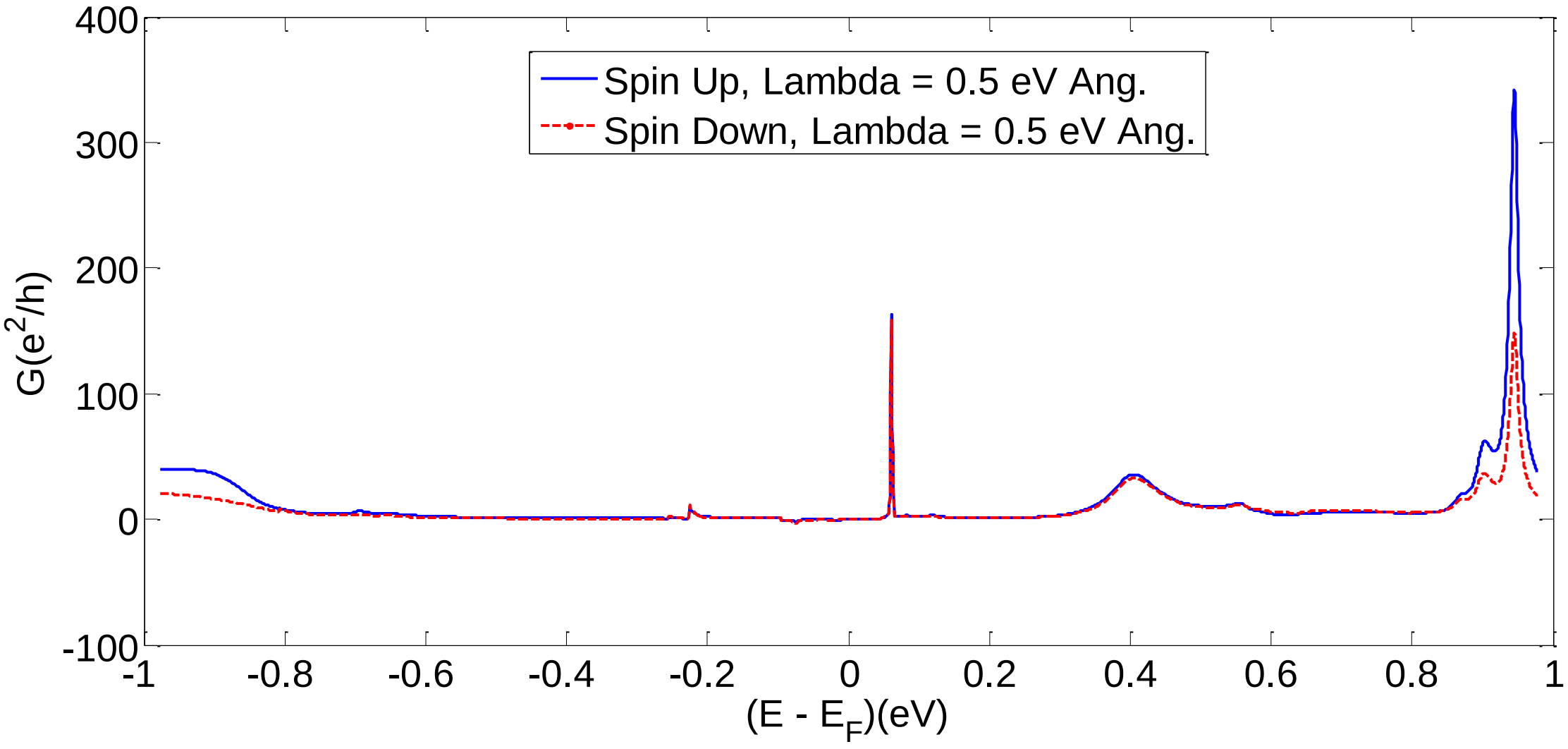

(d)

**Fig.3 (Color online) Conductance versus electron energy when (a) λ=0.0 eV, (b) λ=0.065 eV Ang., (c) λ=0.13 eV, and (d) λ=0.5 eV**

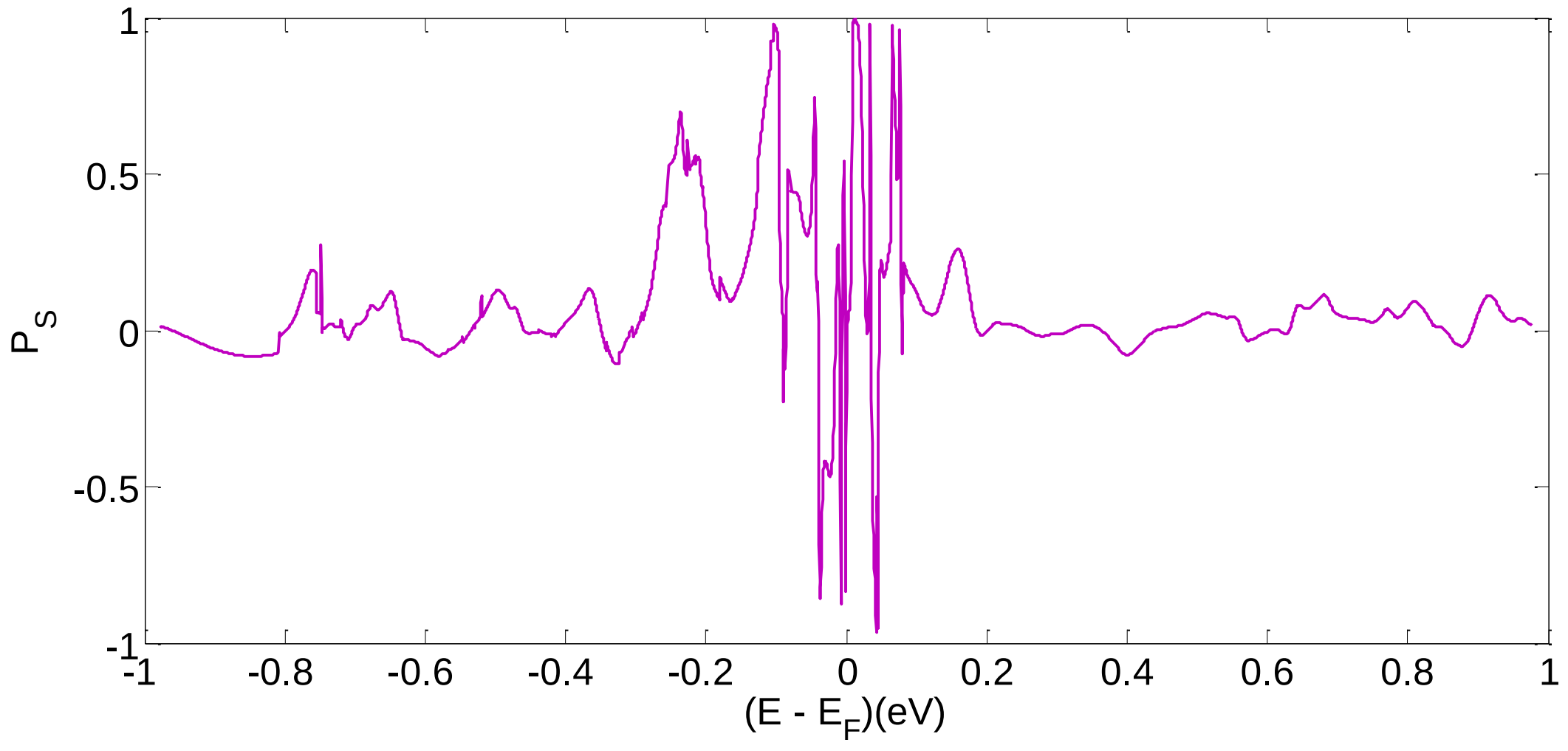

(a)

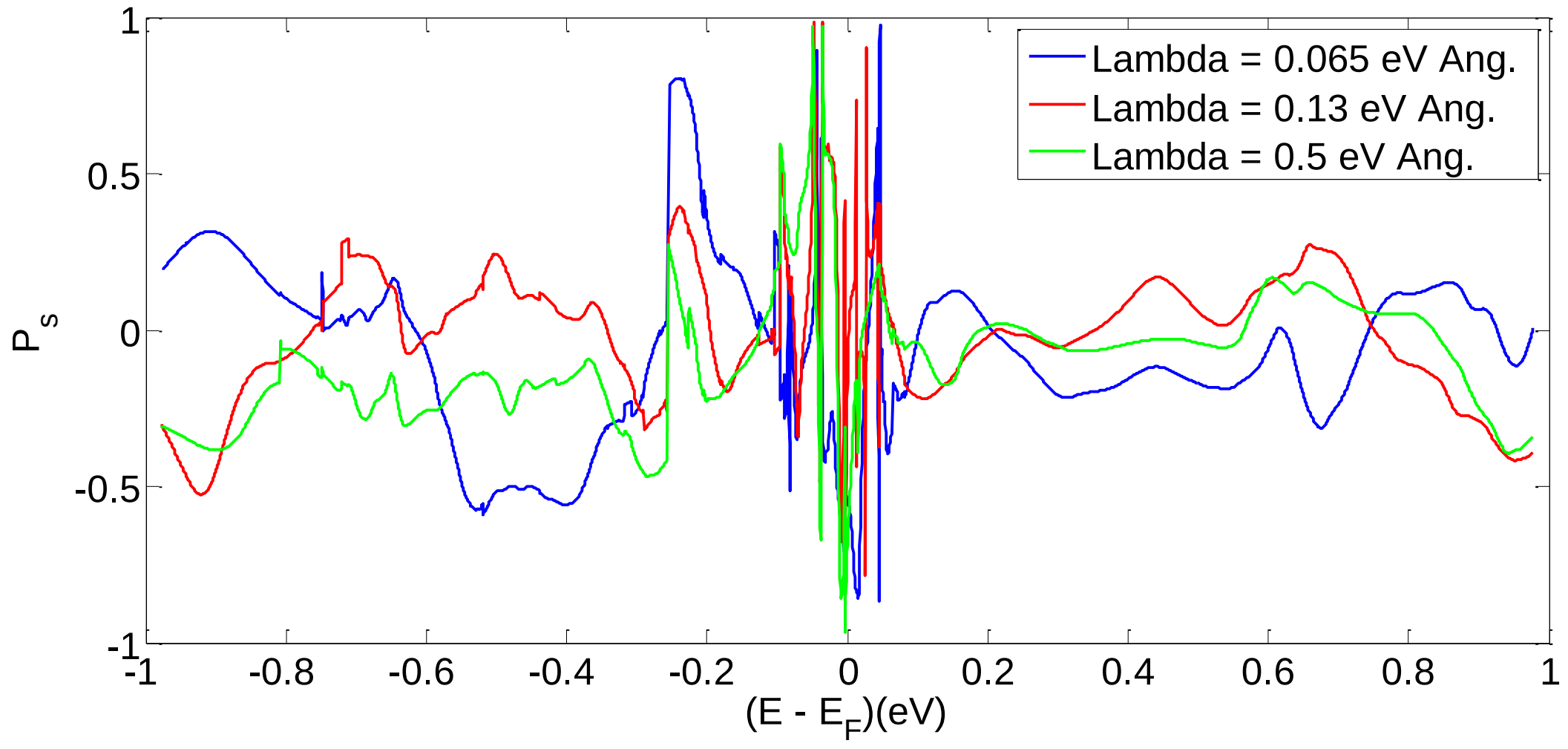

(b)

**Fig. 4 (Color online) Spin polarization versus electron energy (a) without Rashba SOI and (b) with Rashba SOI**